# Utilization of Pulse Rate Variability between Post-sleep and Wake Cycles to increase Alarm Clock Efficiency using Arduino-based Non-invasive Pulse Detector

Daniel D. Cabrales, Jhon Rafael M. Cuartero, Lejan Alfred C. Enriquez, John Gabriel Z. Erne, Bryan Dayton Edward N. Galvadores, Mark Lester S. Millan, Beryl Keziah C. Monterona, John Vincent P. Panergalin, Melbert Neil G. Teodoro,
Lean Karlo S. Tolentino[1]
Electronics Engineering Department, Technological University of the Philippines
Manila, Philippines
Corresponding author: [1] leankarlo.tolentino.ph@ieee.org

*Abstract. The most common reason for students' tardiness is due to waking up late every morning. The causes may vary from lack of rest and sleep, to failure of ascertaining an alarm clock's noise or sometimes, falling back to sleep again after snoozing an alarm. This paper's purpose is to address the problem concerning student's inability to continue to stay awake after turning off their alarms using Arduino-based pulse rate triggered alarm interrupt (APRTAI). The proposed system will enable students to make use of a higher pulse rate from their thumb to turn off an alarm which can be achieved by engaging themselves in intensive physical activities that will eventually eradicate drowsiness and therefore prevent them from falling back to sleep again.*

**Introduction**

This project implements the use of an Arduino compatible board, modified commercial digital LED alarm clock and non-invasive optical pulse monitor. The hardware system is implemented using the pulse rate of one's person using the pulse sensor, while its output signal from the clock is processed by the Arduino compatible board for an output of sound from a source.

Based on a study (Kräuchi and Wirz-Justice, 2001; Pickut, 2013), the pulse rate of a healthy person while asleep decreases by a rate of 8 to 10% from the normal pulse rate which ranges from 80 to 100 beats per minute. The study testifies the fact that pulse rate differs when awake or asleep. On another study (British Heart Foundation, 2014), the pulse rate of a person that is exercising is 50-69% of the maximum heart rate at a moderate degree of exercise. The maximum heart rate value is computed from the difference between 220 and the age of the person (Fox and Haskell, 1970; Tanaka et al., 2001).

Since students tend to go back to sleep after turning the alarm clock this project aims to solve the problem of constantly snoozing the alarm as a result of them not waking up on time and being late for any appointment for the day. The mechanics behind the proposed project is to make use of a student's heightened pulse rate to turn off the alarm which is set at the digital LED alarm clock itself. It aims to force the individual to carry out physical activities first which will induce an increase in heart rate which can be detected by the optical pulse monitor which also in turn, remove the feeling of sleepiness.

The proposed system employs non-invasive method of measuring the pulse rate of an individual. Studies (Kräuchi & Wirz-Justice, 2001; Pickut, 2013; British Heart Foundation, 2014) suggest high positive correlation between the wakefulness and pulse rate of a person. This project utilized this fact to increase the alarm clock's efficiency by forcing the user to undergo physical exercise in order to stop the alarm.

The developed pulse counter code (Heart_Rate_BPM, n.d.) for the optical heart monitor utilized digitally triggered basic control statements. This, however, presents high susceptibility to stray pulses and may ignore pulses with amplitudes between digitally high and digitally low signals (Kuphaldt, 2007) This project employed Schmitt trigger (Poole, 2013) implemented using control statements to resolve the encountered problem.

**Methods**

System design and implementation

*Software design.*

The program focuses mainly on acquiring reliable heart pulse signal from the users' fingertip using the optical heart rate monitor. This is accomplished using Schmitt trigger characteristics implemented using Arduino IDE programming.

The control system of this project, implemented using Arduino IDE programming, is described by the flow chart shown in Figure 1.

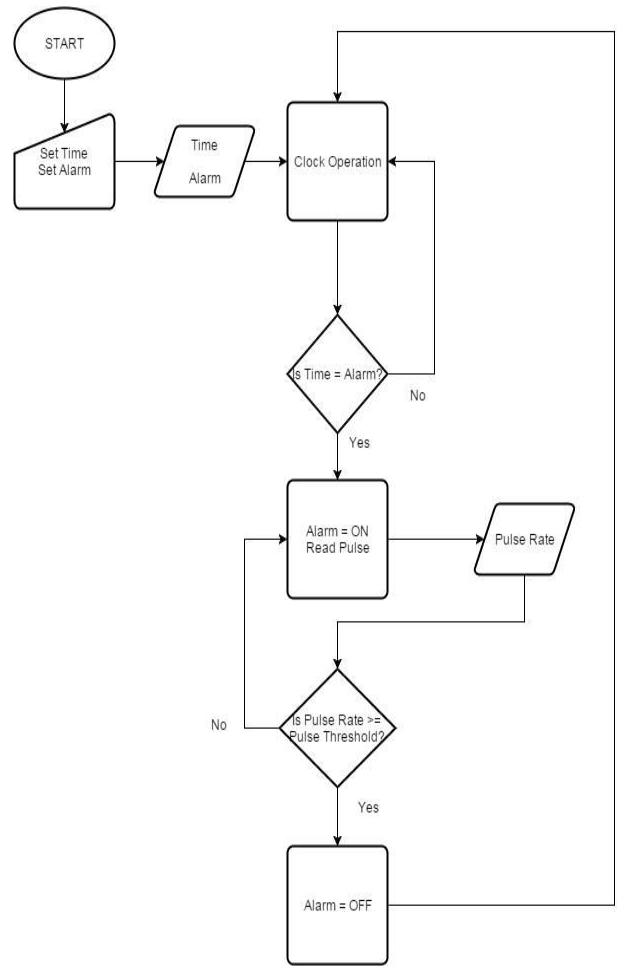

**Figure 1**. Flow chart of the proposed system

*Hardware design.*

The main device of the project is the gizDuino v.5 board as seen in Figure 2, an Arduino compatible board that reads the signal given by the heart rate sensor and controls the system. Added also to the system is a shield (Figure 3) that converts the analog signal from the modified LED clock (Figure 4) alarm system into a digital signal which can be processed easily by the gizDuino board. The output of the system is then fed to a basic 5-volt DC buzzer.

The heart rate sensor used, shown in figure 5, is a simple optical sensor that is compatible to all Arduino-Compatible MCU boards which is equipped with a slot for a fingertip that is ready to read an individual heart rate (E-gizmo, 2015)

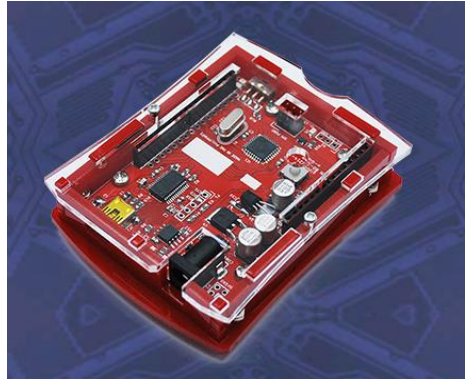

**Figure 2**. gizDuino v.5 board

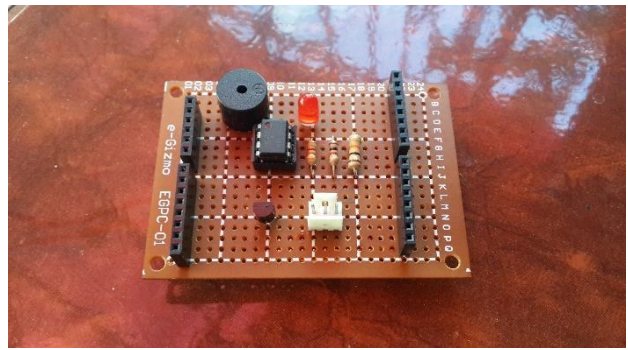

**Figure 3**. gizDuino v.5 board

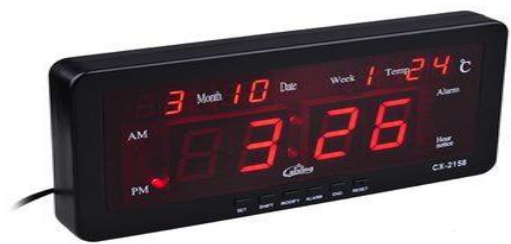

**Figure 4**. Digital LED alarm clock

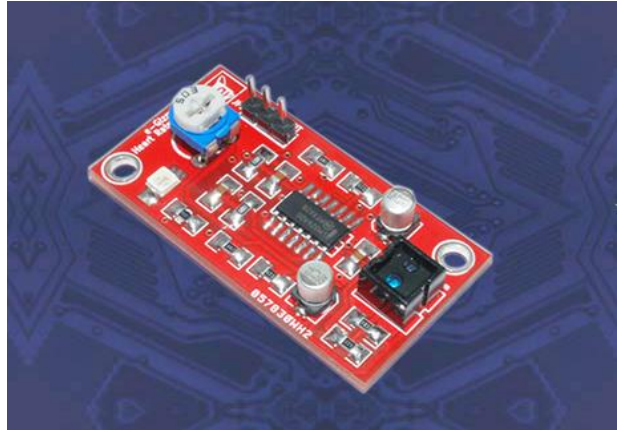

**Figure 5**. Heart rate sensor

Device operation

The purpose of the device is to use the elevated heart rate of the user after a physical exertion which implies wakefulness to effectively awaken him and otherwise, prevent the user from falling back to sleep again.

The procedure starts by setting the alarm clock to a desired time. When the time set is reached, the LED clock sends a signal into the driver shield which is then converted to a digital signal using bistable multivibrator implemented using LM555 mixed signal IC. The digital signal prompts the gizDuino to set off the alarm, activating the buzzer that will continuously run even the time on the clock has already passed the desired alarm time unless an input coming from the heart rate sensor is introduced.

In order to stop the alarm, the users' fingertip must be placed on the allotted slot in the sensor for it to read their heightened heart rate. When pulse rate reading has reached the required heart rate in beats per minute (bpm), the gizDuino will interrupt the signal to the buzzer causing it to stop.

**Results and discussion**

This section described the testing of the project due to its capability and effectiveness. As the alarm was triggered the signal was sent to the gizDuino board for processing. It filters the output that negates the pulse rate reading less than 23 and more than 200 bpm. As seen from

Figure 6 which illustrates the pulse rate readings from the Serial Monitor of the Arduino IDE software, the alarm will momentarily stop as it reaches the pulse rate reading of 101-199 bpm.

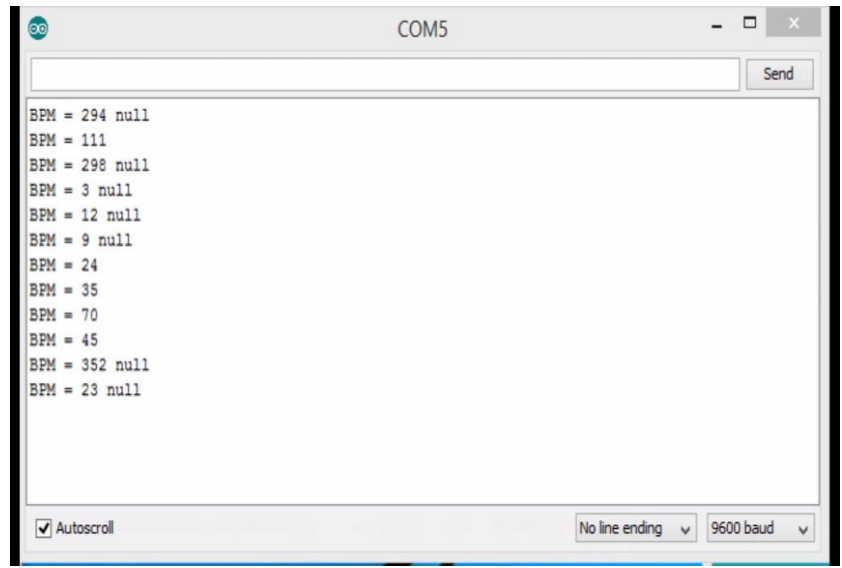

**Figure 6**. Pulse rate readings seen from Arduino serial monitor

**Conclusion and recommendation**

The Arduino-based pulse rate triggered alarm interrupt (APRTAI) is an effective device to keep students from being late from classes or any social appointments by enabling them to wake up on time and prevent them from falling back to sleep again. Unlike any other alarm clocks which can be snoozed or turned off whenever a person wants, the APRTAI forces him to undergo physical activity for benefit of both turning off the annoying sound of the alarm and eradicate the feeling of drowsiness. Furthermore, it can encourage the development of healthy lifestyle and physique as the necessary exercise can directly affect the body.

Further researches are needed to improve the proposed project. A suitable algorithm for the stability of the pulse rate readings will be implemented. Future work may include applications on mobile phone clock alarms through Android or iOS apps, implementation on digital clock watches, or integration of the proposed system on exercise machines and vehicles.

**Acknowledgments**

The authors would like to thank the people who supported them to make this paper possible: their parents, for financial support, and the faculty of the Electronics Engineering Department, College of Engineering of the Technological University of the Philippines – Manila, for imparting invaluable knowledge and techniques in programming and research.